\documentclass[aps,prl,floatfix,twocolumn,superscriptaddress]{revtex4-1}
\usepackage[final]{graphicx}
\usepackage{times,bbm,amsmath,amssymb}
\usepackage{epsfig,color}
\usepackage{hyperref}
\usepackage{float,siunitx}
\usepackage[caption = false]{subfig}
\usepackage[english]{babel}
\usepackage{thumbpdf,enumerate}
\usepackage{booktabs}
\usepackage{sidecap}
\usepackage[scaled=.8]{couriers}
\usepackage{pstricks}
\usepackage{multirow}
\usepackage{placeins}
\usepackage{pst-grad}
\usepackage{epigraph}
\usepackage{longtable}
\usepackage{booktabs}
\usepackage{gensymb}

\begin{document}

\title{Robust spectral phase reconstruction of time-frequency entangled bi-photon states}
%\title{Experimental entanglement-enhanced work extraction}
%\title{Experimental entanglement-enhanced work extraction based on a Maxwell's demon}

\author{Ilaria Gianani}\email{ilaria.gianani@uniroma3.it}
\affiliation{Dipartimento di Scienze, Universit\`a degli Studi Roma Tre, Via della Vasca Navale 84, 00146, Rome, Italy}

\begin{abstract}
Exploitation of time-frequency properties of SPDC photon pairs has recently found application in many endeavors. Complete characterization and control over the states in this degree of freedom is of paramount importance for the development of optical quantum technologies. This is achieved by accessing information both on the joint spectral amplitude and the joint spectral phase. Here we propose a novel scheme based on the MICE algorithm, which aims at reconstructing the joint spectral phase by adopting a multi-shear approach, making the technique suitable for noisy environments.
We report on simulations for the phase reconstruction and propose an experiment using a Franson modified interferometer.

\end{abstract}

\maketitle

%Time frequency utile 
%
%classicamente faccio multiplexing
%
%Per usare questa tecnologia ccorre sapere cosa faccio
%
%Normalmente per caratterizzare faccio quantum state tomography 
%
%Però non è praticabile col time-frequency quindi vai a vedere se le risorse classiche ti aiutano
%
%Classicamente tutto parte da spider che è poi stato complicato a piacere per vedere impulsi più complessi --> Multishear robustezza. MICE!
%
%MICE usa la ridondaza per garantire una ricostruzione anche con alti livelli di rumore. 

\begin{figure}[h]
\includegraphics[width=0.9\linewidth]{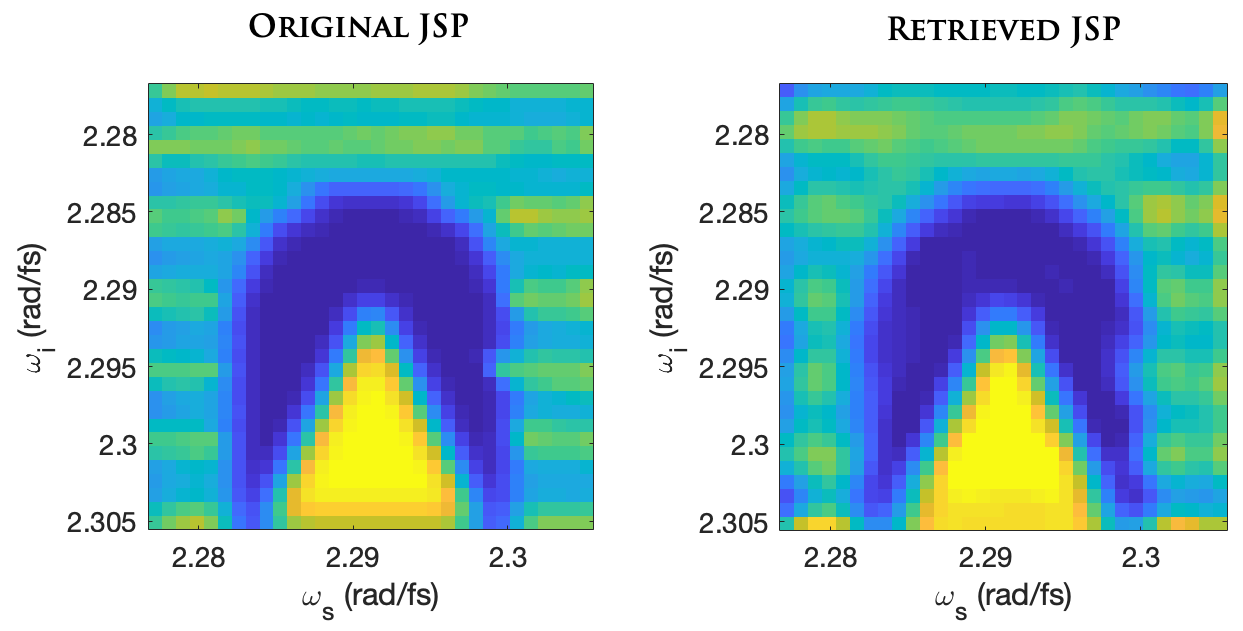}
\caption{{\it JSP reconstruction} of University of Roma Tre logo.}
\label{rm3}
\end{figure}

Spectral-temporal properties are amongst the most reliable and robust choices for encoding information for photonics quantum techonolgies. Being an internal degree of freedom, it  is suitable for long distance communications, and allows for propagation through long-distance fibres without affecting the quantum state \cite{fibreprop}. Applications exploiting time-frequency encoding range from QKD protocols \cite{josh, weiner, rodiger, mower}, clock synchronization \cite{giov}, and quantum communications \cite{wiseman}, all of which make use of frequenncy correlated two-photon states. \\
\indent The most common technique to generate such pairs are non linear processes, such as spontaneous parametric down conversion (SPDC) and four wave mixing, in which the spectral-temporal properties are ditcated by the shape of the pump as well as the material through its phase-matching function. Tailoring the pump and choosing the appropriate dispersion grants for diverse capacities in shaping spectrally broad two-photon states \cite{vahid, raymer, mosley, carrasco}.  
Quantum technologies demand that the information carriers are prepared in fiducial states at the beginning, as a key requirement for the correct operation of any protocol. The variegate structure of time-frequency modes is at the same time what grants its advantages but poses some critical challenges in its characterization. Ultrafast pulsed modes - exactly like their classical counterpart - vary too quickly to be characterized in the time domain. To characterize them in the frequency domain, there is need to access both their spectral amplitude and spectral phase, as both affect the time profile and can carry signatures of frequency correlations. Measuring the joint spectral amplitude is now a commonly addressed task \cite{brian2,xstine,clark}, however the measurement of the joint spectral phase has only recently been tackled. This has been achieved by performing quantum state tomography on the biphoton state \cite{john}, by extending what is normally applied to discrete systems, e.g. polarization. \\
\indent An alternative route relies on classical ultrafast metrology techniques, which have been extensively developed following the need to characterize femtosecond and attosecond pulses \cite{ian}. An approach in this direction has been recently proposed in \cite{brian}, where the self- reference classical metrology technique SPIDER \cite{spider} has been adapted to the heralded measurement of photon pairs phases.  SPIDER reconstructs the spectral phase by retrieving the interferometric phase between two frequency-sheared copies of an unknown pulse. The extraction algorithm is quite simple and is based on the integration of the interferometric phase. In the last few decades many different implementations of SPIDER have been developed to address increasing degrees of pulse complexity \cite{spider2, spider3,spider4,spider5}. In particular multi-shear techniques as SEA-CAR SPIDER \cite{carspider1, carspider2,carspider3} have provided a very robust tool for the reconstruction of broadband pulses with high spectral complexity. In all its implementations SPIDER is a referenced technique, where the reference is either the pulse itself, or, with a slight modification, a known external field. More recently a new algorithm, MICE \cite{MICE}, which relies as well on multi-shear techniques, has been developed. Contrary to the standard SPIDER extraction algorithm, MICE allows for the mutual characterisation of multiple unknown fields at the same time. Due to the redundancies achieved via the multi-shear arrangement, MICE performs extremely well even under very stringent noise conditions.  This technique has proven to be extremely versatile in the classical regime and it has been employed for the reconstruction of spectral phases of complex pulses in the visible-near IR regime \cite{spice}, for wavefront reconstruction \cite{MICE}, for the spatial characterization of high harmonic sources \cite{mmm}, and for digital holography microscopy \cite{patrick}. \\
\indent Here we propose a technique to employ MICE in a setup taking into accout the specific needs of quantum light detection. The measurement strategy relies on the use of a modified Franson interferometer  \cite{cabello09, hardy11}, which allows to observe genuine time-bin entanglment without relying on time-resolved detection. This is a necessary condition to obtain coincidences which are dependent on the biphoton spectral phase to be extracted. Simulations show that, due to the redundancy provided by the multishear approach, the technique  works reliably even with moderate signal intensities.\\
\indent MICE is a classical metrology technique which uses an iterative algorithm to simultaneously reconstruct multiple unknown fields $E_i$ depending on a set of parameters $\gamma$ without the need of an external known reference. This is made possible by means of a multishear measurement strategy, in which multiple shears must be used to scan the fields along each parameter, and the number of fields to reconstruct mus be lower than the number of shears used for each dimension. This is sufficient to guarantee enough redundancy, which makes the technique particularly robust against noise. 
Given two fields $E_1(\gamma)$ and $E_2(\gamma)$, MICE relies on the minimization of the error with respect to each field \cite{MICE}: 
\begin{equation}
\mathcal{E}= \sum_{j,k,l} \vert AC^{meas}_{j,j-k} - E_1(\gamma_j)E_2^*(\gamma_j-\Gamma_{k}) \vert ^2
\label{error}
\end{equation}
where $AC^{meas} $, is the measured interferometric product between the two fields, obtained as the sideband of the Fourier transform of $I=\vert E_1(\gamma) + E_2(\gamma-\Gamma_{k}) \vert ^2$. Measuring the bi-photon spectral phase requires implementing interferometric schemes, which typically demand for long accumulation times to achieve good signal levels. Given its robustness against noise, using MICE grants a solution to this, becoming the preferable choice for such an endeavor. This is conditioned on properly choosing an arrangement whose measurement outcome obey to the behavior described above.
\begin{figure}[h]
\centering
\includegraphics[width=1\linewidth]{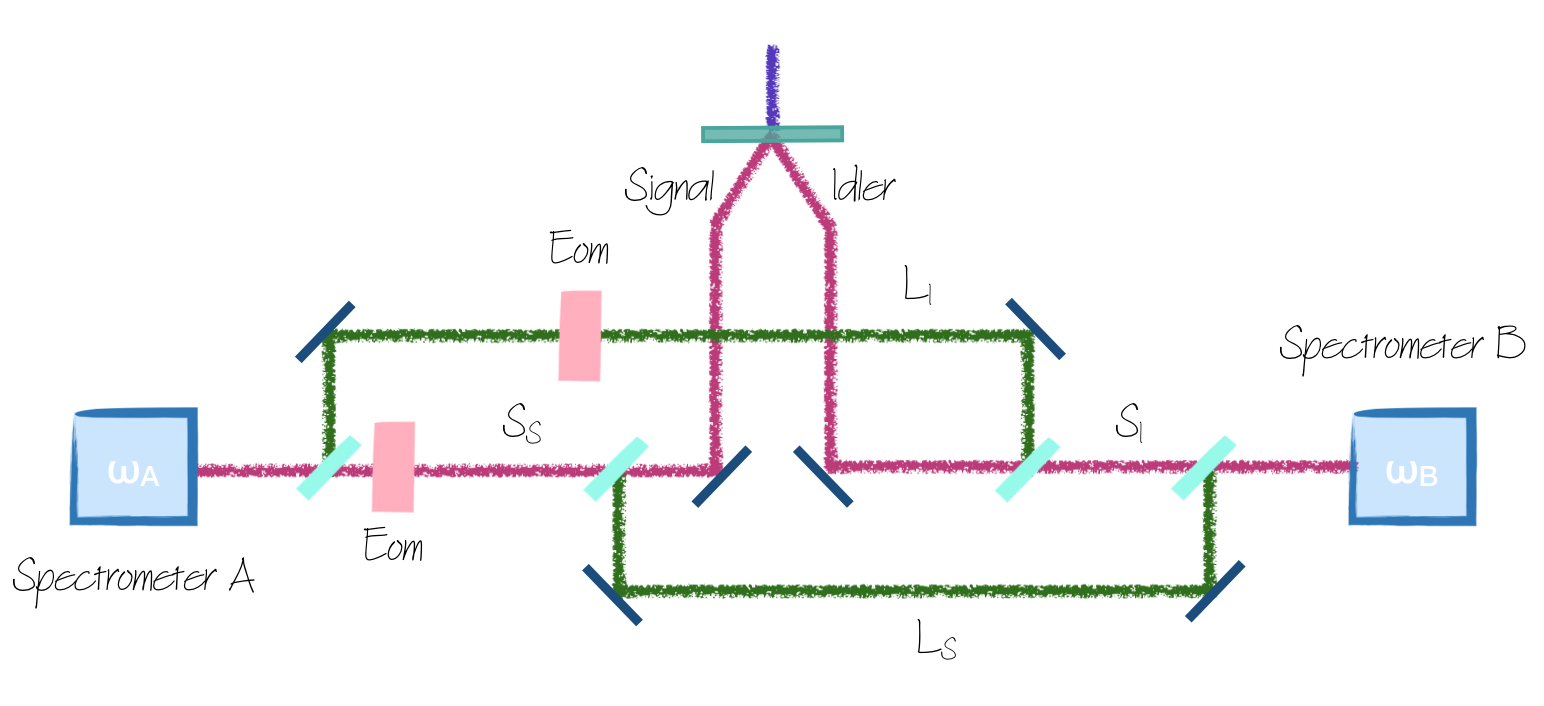}
\caption{{\it Proposed interferometric scheme.} A photon-pair produced via SPDC enters a modified Franson interferometer where each photon can undertake either a long (L) or short (S) path. When the signal  (idler) photon passes through the short (long) path it is subject to a frequency shear given by the EOM. A frequency-resolved coincidence counting measurement is then performed.}
\label{setup}
\end{figure}
Consider the modified version of the interferometric scheme proposed by Cabello et. al \cite{cabello09}, as depicted in Fig. \ref{setup}. The original motivation of this scheme lies in easing some technical requirements of Franson's original idea \cite{franson} for the generation of time-bin entanglement. A photon pair is generated via spontaneous parametric down conversion(SPDC); both the signal and idler photons can undertake either a short $\vert S \rangle$ or a long $\vert L \rangle$ path before being detected with a frequency-resolved measurement. This scheme has been proved to generate time-bin entanglement between the short and long paths without relying on time-resolved detection \cite{cabello09} . In order to make it suitable for our purposes, two further modifications need to be introduced: frequency resolved detection is adopted; independent frequency shears are inserted on the $S_s$  and $L_i$ path: the signal will be sheared only when taking the short path, the idler only when taking the long one. Shearing can be performed by means of Electro Optic Modulators (EOMs) as proposed and demonstrated by \cite{eosi, brian}. This is preferable to non-linear optical shearing, as we work in the single photon regime. Since we adopt a multishear approach, both the shears have to be scanned independently through multiple values, so that for each shear $(\Omega_{1,k})$ on  $S_s$, the shear $(\Omega_{2,l})$  on $L_i$ scans along the idler dimension of the joint spectral wavefunction. In the most general case, MICE is not bounded to work with fields having the same spectral support, if the shears are chosen so that the interferograms will completely cover the fields along every dimension. In fact, the phase will be only retrieved in the zones covered by the interference. At the same time the interferograms given by two subsequent shears need to overlap, otherwise it cannot make use of the redundancy. If the fields interfering have the same isupport, the sole purpose of the multiple shears is to grant the redundancy, hence they can be as small as allowed by the detection resolution. The state entering the interferometer will be given by \cite{eberly}:
\begin{equation}
\Psi(\omega_s,\omega_i)=\int d\omega_s \,d\omega_i A(\omega_s,\omega_i)a^{\dagger}_s(\omega_s)a^{\dagger}_i(\omega_i)\vert 0 \rangle\vert 0 \rangle,
\end{equation} 
where $ A(\omega_s,\omega_i)$ is the wave function of the  biphoton state. As the photon pairs goes through the interferometer, the output state will be transformed into:
\begin{equation}
\begin{aligned}
\Psi(\omega_s,\omega_i) = \int d\omega_s \,d\omega_i A(\omega_s,\omega_i)a^{\dagger}_s(\omega_s+\Omega_{1,k})a^{\dagger}_i(\omega_i)+\\
+A(\omega_s,\omega_i)a^{\dagger}_s(\omega_s)a^{\dagger}_i(\omega_i+\Omega_{2,l})e^{i(\omega_s+\omega_i+\Omega_{2,l})\tau}\vert 0 \rangle\vert 0 \rangle,
\end{aligned}
\end{equation}
where $\tau$ is the delay between the two paths and we have supposed to perform the shear on the $L_i$ path after a length equal to that of the S paths. We notice however that due to the modified geometry, the detector will not always measure $\omega_s$ or $\omega_i$, and that is a fundamental requirement to assure genuine time-bin entanglement between the two photons, as it allows to automatically discard the $\vert S \rangle \vert L \rangle$ ad  $\vert L \rangle \vert S \rangle$ events. 
Hence, when the photons are measured, the destruction operators, as a function of the measured frequencies $\omega_A$ and $\omega_B$, are $ b_A(\omega_A) = a_s(\omega),\,\,\, b_B(\omega_B)=a_i(\omega) $, if the state measured is $\vert SS\rangle$, or 
$b_A(\omega_A) = a_i(\omega),\,\,\, b_B(\omega_B)=a_s(\omega)$ if the state measured is $\vert LL \rangle$. Hence, the coincidence probability reads:
\begin{equation}
\begin{aligned}
&P(\omega_A,\omega_B)=\\
&\vert A(\omega_A - \Omega_1,\omega_B) + A(\omega_B, \omega_A - \Omega_2)e^{i(\omega_A+\omega_B)\tau} \vert^2 
\end{aligned}
\end{equation}
We can now define $E_1(\omega_s,\omega_i) \equiv A(\omega_A - \Omega_1,\omega_B) = X_1(\omega_A - \Omega_1,\omega_B)e^{i\phi_1((\omega_A - \Omega_1,\omega_B))}$ and $E_2(\omega_s,\omega_i) \equiv A(\omega_B, \omega_A - \Omega_2) = X_2(\omega_B, \omega_A - \Omega_2) e^{i\left[\phi_2(\omega_B, \omega_A - \Omega_2)+(\omega_A+\omega_B)\tau\right]},$where $X_i$ is the spectral amplitude of each electric field ad $\phi_i$ is its phase, ad we have introduced the ancillary field $E_2(\omega_s,\omega_i)$, which bears no additional physical meaning but is instrumental to the field retrieval. The probability becomes:
\begin{equation}
P(\omega_A,\omega_B)=\vert E_1(\omega_A - \Omega_1,\omega_B) + E_2(\omega_B,\omega_A-\Omega_2) \vert^2,
\end{equation}
which has the same structure of the interferogram $I$ between $E_1(\gamma)$ and $E_2(\gamma)$, where $\gamma = \omega_s,\omega_i$. As such, this can now be processed with the MICE algorithm, solving the following equations which have been obtained by minimizing the error in Eq. \eqref{error} with respect to both fields:
\begin{equation}
\begin{aligned}
&E_1(\omega_i,\omega_j)=\frac{\sum_{k,l} AC^{meas}_{i -k,j+l}\cdot E_2^*(\omega_{i}+\Omega_{1,k},\omega_{j}-\Omega_{2,l})}{\sum_{k,l}E_2(\omega_{i}-\Omega_{1,k},\omega_{j}+\Omega_{2,l})}\\
&E_2^*(\omega_i,\omega_j)=\frac{\sum_{k,l}AC^{meas}_{i+k,j-l} \cdot E_1^*(\omega_{i}-\Omega_{1,k},\omega_{j}+\Omega_{2,l}) }{\sum_{k,l}E_1(\omega_{i}-\Omega_{1,k},\omega_{j}+\Omega_{2,l})}.
\end{aligned}
\end{equation}
To solve this set of equations it is necessary to provide an initial guess for $E_2$, so to obtain an initial value of $E_1$ which is then fed into the second equation. By iteration the two fields are retrieved. We remark that as any implementation based on SPIDER, MICE suffers form  ambiguities in determining the amplitude $X(\omega_s,\omega_i)$: the phases retrieved for both fields will be accurate, but the amplitudes will not. In order to retrieve the JSA with the setup proposed, it would be sufficient to block the L arms, and perform the spectral measurement on the S arms alone. 
\begin{figure}[h]
\centering
\includegraphics[width=1\linewidth]{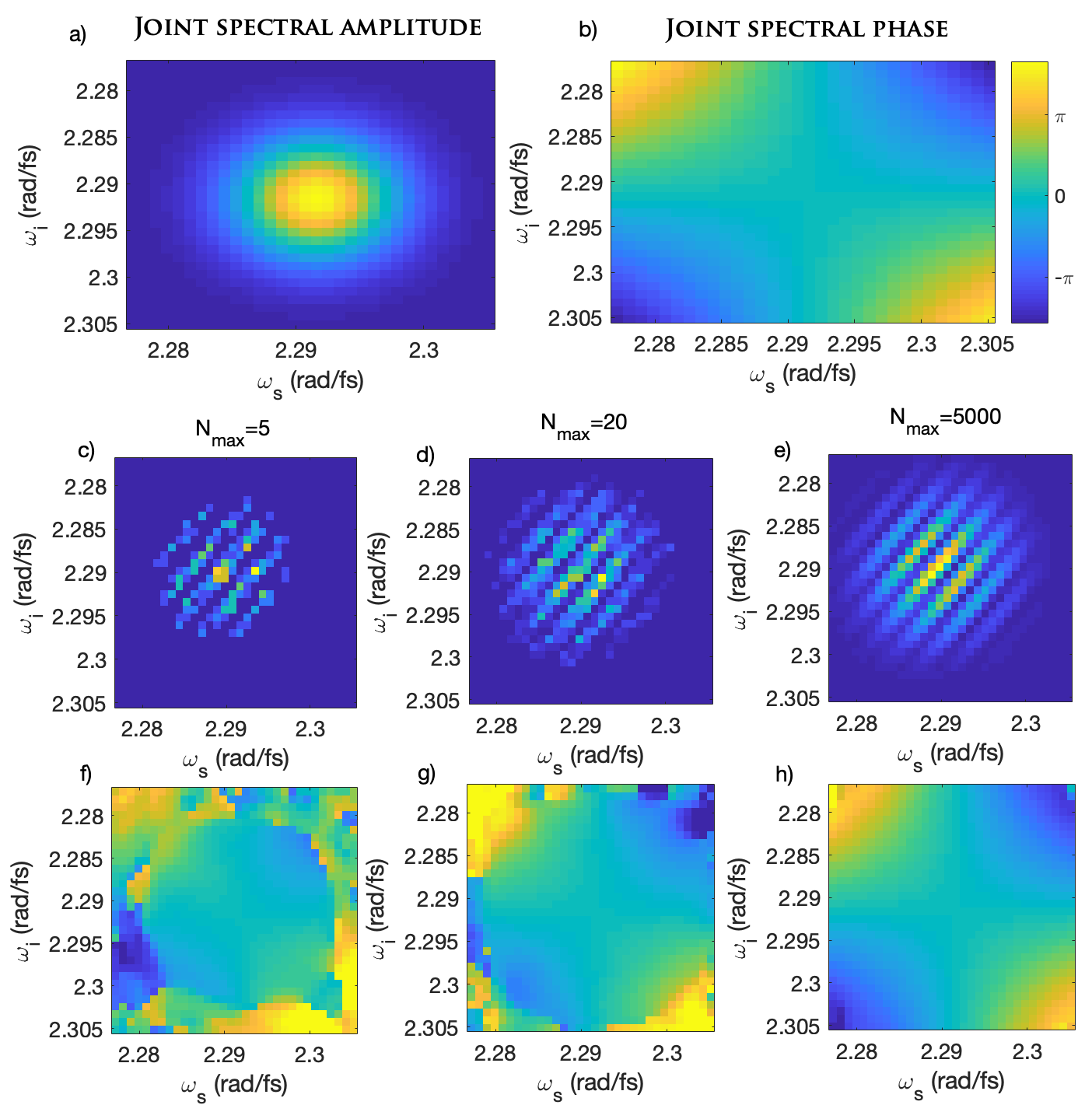}
\caption{{\it Simulation restults.} a) Joint spectral amplitude for $E_1$ b) Joint spectral phase for $E_1$ c-e) interferograms obtained with 5, 20, 5000 maximum peak coincidence counts f-h) retrieved joint spectral phase for the three signal intensities}
\label{results}
\end{figure}
In order to test the analysis routine, we perform a bi-photon phase reconstruction on simulated data. The JSA and JSP constituting $E_1$ are shown in panel (a) and (b) of Fig. \ref{results}, and are those emitted by typical (e.g. those in \cite{brian, john}). The field is sampled on a 32x32 pixels grid, covering a spectral range of 10 nm centered at 820 nm along each dimension. $E_2$ is obtained as a permutation between the two dimensions of $E_1$. The shears are then applied to both fields. Both $\Omega_1$ and $\Omega_2$ can each assume 8 different values, which leads to 64 interferograms per reconstruction. As per Fig. \ref{results}, we choose a scenario in which the correlations are present only in the phase to be retrieved and not in the JSA, which results in both $E_1$ and $E_2$ sharing the same amplitude. Since the amplitude of the two fields is the same and the multiple shears are used only for the required redundancy, their value can be as small as dictated by the detection resolution, so in order to have eight different values, the shear on each arm will vary between -4 px to 3 px (where each pixel corresponds to $\Delta \lambda \sim 0,3$ nm). \\
\indent To test against the robustness to noise in a realistic scenario, we perform different reconstructions by varying the signal intensity. In particular the interferograms are normalized by setting peak coincidence counts of the interferogram $N_{max} $, from 5 to 8000 coincidences. Furthermore accidental coincidences are added accordingly, given by $N_{acc}=(N_{max}/0.1)^2/80e6$, obtained considering a $10\%$ coincidence efficiency to determine the signal intensity and a repetition rate of 80 MHz. Note that $N_{acc}$ is calculated on the maximum coincidence value and is hence overestimated. The interferograms are then randomly generated with a Poissonian distribution centered at the value give by the normalization for each pixel, to which Possionian-distributed accidental coincidences are added. The interferograms are then fed to the MICE algorithm set with 20 iterations. The interferograms for $N_{max} = 5 cc$, $N_{max} =20 cc$, and $N_{max} =5000 cc$ are shown in Fig. \ref{results} panels c-e.  Panels f-h show the reconstructed JSP of $E_1$ for each signal intensity. The phase of $E_2$, which is also reconstructed, is not shown as it does not add any meaningful information. 
\begin{figure}[t]
\includegraphics[width=0.9\linewidth]{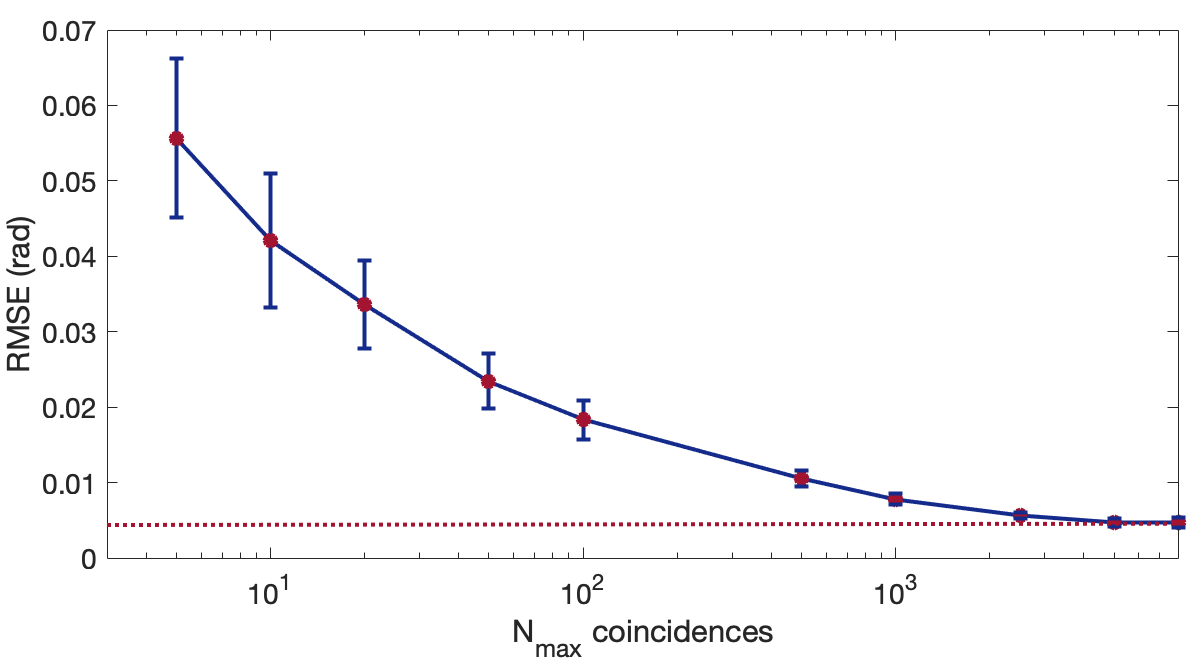}
\caption{{\it RMS error} between the original and retrieved phase vs. the interferogram's peak coincidence counts.The error saturates at 5000 peak coincidences at $4.5\cdot 10^{-3}$ rad .} \label{rmse}
\end{figure}
Each reconstruction is then repeated 30 times to accumulate statistics for calculating the RMS error, weighted with the field's intensity  \cite{articolodiian}, between the original and retrieved phase of $E_1$. The results are shown in Fig. \ref{rmse}.  Varying the signal intensity, the error converges to its minimum of $0,0045$ rad for $N_{max} = 5000$.  However even for 5 peak counts the intensity-weighted RMSE is 0.056 rad, which indicates a good agreement between the retrieved and original phase. In fact, even when the full span of the phase is not reconstructed, the low intensity doesn't affect the reconstruction in the portion with non-zero signal. This makes MICE an excellent tool for dealing with particularly low count rates and noisy scenarios. Shear, resolution, and signal intensity all concur in achieving a correct reconstruction and have to be tailored to the measured state, taking into account its spectral amplitude and phase complexity, which is common to every reconstruction technique in the classical domain as well. Nonetheless, with the appropriate choice of parameters,  the algorithm is capable of successfully reconstructing arbitrarily complex JSPs, as demonstrated on the reconstruction in Fig. \ref{rm3}, where the University of Roma Tre logo has been used as JSP. With respect to the reconstruction shown before, 32 shears were employed instead of 8, however the same resolution and spectral intensity were kept of the previous, more realistic, case. In this example the redundancy given by the multiple shears contrasts the lack of resolution in the reconstruction of a highly structured phase, showing the flexibility given by the interplay among the many reconstruction parameters.
\\
\indent Concluding, we propose of a new technique which is capable of characterizing the joint spectral phase of a biphoton state even in low-signal, noisy regimes. This takes advantage of the high redundancy granted by the multi-shear approach, which is implemented using a modified Franson interferometer. The robustness to noise is reflected in a rapid convergence of the RMS error to its minimum.  The proposed setup presents its complexities but it has already been successfully used in many endeavours. The lack of strict signal requirements and the robustness to noise make up for these complexities, posing this novel technique as possible route to obtain a complete characterization of time-frequency states. \\
{\it Acknowledgements.} The author would like to thank M. Barbieri for his helpful advice, and G. Vallone for fruitful discussion.
%
%\begin{itemize}
%\item Interest in characterising the JSP of a photon source  - provide a diagnostic tool for SPDC quantum technologies
%\item Done by Brian using SPIDER twice (EOSI)
%\item the technique works, however it requires to perform the shear on the two dimension at subsequent times  
%\item propose a technique based on MICE algorithm that uses the redundancy provided by the shear and as such is demonstrated to be extremely robust to noise.
%\item to do so, we consider a franson like setup modified so to add a shear on S and L
%\item probability of having a coincidence between the S and L path is...
%\item this can be seen as interference between E1 and E2 and fed to  MICE 
%\item MICE equations \& simulations
%
%\end{itemize}

\end{document}